\documentclass[doublecol]{epl2} 

\usepackage{graphicx}
\usepackage{bm}
\usepackage{bbold}
\usepackage{color}

\newcommand{\bea}{\begin{eqnarray}}

\newcommand{\eea}{\end{eqnarray}}

\newcommand{\beq}{\begin{equation}}

\newcommand{\eeq}{\end{equation}}

\title{Electron-electron interactions in graphene bilayers}

\author{S.~Viola~Kusminskiy \inst{1} \and D.~K.~Campbell \inst{1} \and A.~H.~Castro~Neto \inst{1}}

\institute{                    
  \inst{1} Department of Physics, Boston University, 590 Commonwealth Ave., Boston, MA 02215}

\abstract{
We study the effect of electron-electron interactions in the quasiparticle
dispersion of a graphene bilayer within the Hartree-Fock-Thomas-Fermi theory by using a four-bands model.
We find that the electronic fluid can be described by a non-interacting like dispersion but with renormalized parameters.  We compare our results with recent cyclotron resonance experiments in this system.}


\begin{document}

\maketitle

\def \tr{{\mbox{tr~}}}
\def \rga{{\rgightarrow}}
\def \ua{{\uparrow}}
\def \da{{\downarrow}}
\def \be{\begin{equation}}
\def \ee{\end{equation}}
\def \ba{\begin{array}}
\def \ea{\end{array}}
\def \bea{\begin{eqnarray}}
\def \eea{\end{eqnarray}}
\def \nn{\nonumber}
\def \l{\left}
\def \rg{\right}
\def \half{{\frac{1}{2}}}
\def \etal{{\it {et al}}}
\def \cH{{\cal{H}}}
\def \cE{{\cal{E}}}
\def \cK{{\cal{K}}}
\def \cM{{\cal{M}}}
\def \cN{{\cal{N}}}
\def \cQ{{\cal Q}}
\def \cI{{\cal I}}
\def \cV{{\cal V}}
\def \cG{{\cal G}}
\def \cF{{\cal F}}
\def \cZ{{\cal Z}}
\def \cC{{\cal C}}
\def \cO{{\cal O}}
\def \bS{{\bf S}}
\def \bI{{\bf I}}
\def \bL{{\bf L}}
\def \bG{{\bf G}}
\def \bQ{{\bf Q}}
\def \bK{{\bf K}}
\def \bR{{\bf R}}
\def \br{{\bf r}}
\def \bu{{\bf u}}
\def \bp{{\bf p}}
\def \bq{{\bf q}}
\def \bk{{\bf k}}
\def \bz{{\bf z}}
\def \bx{{\bf x}}
\def \bpsi{{\bar{\psi}}}
\def \tJ{{\tilde{J}}}
\def \W{{\Omega}}
\def \e{{\epsilon}}
\def \lam{{\lambda}}
\def \L{{\Lambda}}
\def \a{{\alpha}}
\def \t{{\theta}}
\def \T{{\Theta}}
\def \b{{\beta}}
\def \g{{\gamma}}
\def \D{{\Delta}}
\def \d{{\delta}}
\def \w{{\omega}}
\def \s{{\sigma}}
\def \f{{\phi}}
\def \F{{\Phi}}
\def \x{{\chi}}
\def \e{{\epsilon}}
\def \h{{\eta}}
\def \G{{\Gamma}}
\def \z{{\zeta}}
\def \hatt{{\hat{\t}}}
\def \hn{{\bar{n}}}
\def \vk{{{\bf k}}}
\def \vq{{{\bf q}}}
\def \vp{{{\bf p}}}
\def \gk{{\g_{\vk}}}
\def \nd{{^{\vphantom{\dagger}}}}
\def \yd{^\dagger}
\def \av#1{{\langle#1\rgangle}}
\def \ket#1{{\,|\,#1\,\rgangle\,}}
\def \bra#1{{\,\langle\,#1\,|\,}}
\def \braket#1#2{{\,\langle\,#1\,|\,#2\,\rgangle\,}}

Since graphene was isolated in 2004 \cite{Novoselov04}, it has attracted attention
because of its possible application in all-carbon based electronic devices \cite{geim_review}
and its connections to relativistic field theory \cite{pw}. While there is
strong theoretical \cite{rmp} and experimental evidence
\cite{geim_review,yacoby} that single layer graphene (SLG) behaves as essentially a
weakly interacting gas of two-dimensional (2D) Dirac particles,
the situation in bilayer graphene (BLG) is much less clear. Early
theoretical studies have indicated that the SLG is
much less prone towards magnetic states \cite{ferro_single}, while BLG can
become magnetic at low densities \cite{johan_ferro}.
Moreover, while the electronic compressibility of SLG has essentially
features of an insulator \cite{yacoby,macdonald,hwang07}, the
BLG compressibility is, unlike the 2D electron gas (2DEG) \cite{eisenstein}, 
non-monotonic and strongly dependent on electronic density \cite{Silvia08}.
It has also been argued that, unlike SLG, BLG should be unstable towards many-body states such as a pseudospin magnet \cite{PSMmin08}, a Wigner crystal \cite{dahal07}, and an 
excitonic superfluid \cite{min08}. 
It has been demonstrated that BLG is a tunable gap semiconductor 
by application of a transverse electric field \cite{EVCastro07,vandersyppen}, 
leading to extra flexibility in dealing with its
electronic properties \cite{stauber06,ecastro07}.
While electrons in BLG
have a different topological (Berry's) phase than electrons in SLG, as
evident in integer quantum Hall effect measurements 
\cite{novoselov06}, the experimental evidence for electron-electron interaction effects in
BLG has been elusive. Nevertheless, recent cyclotron resonance
experiments in bilayer graphene \cite{Henriksen08} have shown departures from the non-interacting bilayer model proposed
by McCann and Falko \cite{falko}. These disagreements do not seem to be describable in terms of disorder effects alone 
\cite{johan}. The objective of our paper is to clarify these discrepancies.

The SLG has a honeycomb lattice structure that
leads to a Dirac-like electronic dispersion, $E({\bf k}) = \pm \tilde{c} |{\bf
  k}|$, at the edges (the K and K' points) of the Brillouin zone. 
The electrons are described in terms of a 2D ``relativistic'' Dirac
Hamiltonian with zero rest mass, 
where the velocity of light, $c$, is replaced by the Fermi-Dirac velocity, $\tilde{c}$. 
In the BLG (Bernal structure) the two graphene layers are rotated by a
relative angle of $\pi/3$ that
breaks the sublattice symmetry leading to $2$ pairs of massive Dirac
particles at the K (K') point.
Nevertheless, the system remains metallic because $2$ bands, belonging
to different pairs, touch in a point. More explicitly, the non-interacting
bands have the form: $E_{1}({\bf k})=-m \tilde{c}^2+E({\bf k})$, $E_{2}({\bf k})=m
\tilde{c}^2-E({\bf k})$, $E_{3}({\bf k})=m \tilde{c}^2+E({\bf k})$ and $E_{4}({\bf k})=-m
\tilde{c}^2-E({\bf k})$, 
where  $E({\bf k})=\sqrt{(m \tilde{c}^2)^2+ (\tilde{c} {\bf k})^2}$. 
Hence, $E_1({\bf k})$ and $E_4({\bf k})$ ($E_2({\bf k})$ and $E_3({\bf k})$) describe a massive
relativistic dispersion with rest mass energy given by $m \tilde{c}^2$. 
Rotations by other angles do not break the sublattice symmetry
and hence do not lead to mass generation \cite{twist}.  

Our results suggest that BLG behaves as a liquid of Dirac quasiparticles with renormalized
mass and velocity. The situation described here is unique when compared to
standard non-relativistic Fermi liquids such as 
$^3$He \cite{pethick} and ordinary metals \cite{vignale}, or even to
relativistic Fermi liquids such as quark matter in the
core of neutron stars \cite{baym}. While the electrons in
graphene are effectively ``relativistic'', in the 
sense that they obey an {\it effective} Lorentz invariance (only true at low
energies) with the Dirac velocity 
playing the role of velocity of light, on the other hand, from the point of
view of an external observer, the whole graphene system
is Galilean invariant and non-relativistic since the Dirac velocity is much
smaller than the actual speed of light. As a consequence, electron-electron
interactions,  
just as in the case of relativistic \cite{baym} and non-relativistic
\cite{pethick} Fermi liquids, renormalize the quasiparticle mass, but unlike
the relativistic and non-relativistic Fermi liquids, the ``velocity of
light'' is also renormalized \cite{note0}. Moreover, in BLG as well as in SLG the existence of negative energy bands and a pseudo-spin degree of freedom has to be included when constructing a Fermi liquid theory. This has been analyzed microscopically in references \cite{polini07,dassarma07} and through a phenomenological model in Ref. \cite{katsnelson08} for SLG, however a treatment of BLG which takes into account its full hyperbolic four-bands structure is missing.

The Coulomb interaction between electrons breaks the {\it effective} Lorentz invariance of the non-interacting problem 
since it can be thought as instantaneous from the point of view of the
electrons. In SLG, this violation of Lorentz 
invariance leads to the famous upward logarithmic renormalization of the
Fermi-Dirac velocity originally proposed in Ref. \cite{gonzalez}. That effect is a result of the lack
of screening in the SLG due to the vanishing of the density 
of states. The BLG, however, has a finite density of states at the Dirac
point and hence screening plays an important role \cite{hwang08}. 
We show that, similarly to the 2DEG \cite{vignale}, the Hartree-Fock (HF) theory
alone leads to an unphysical logarithmic singularity at the Fermi 
surface indicating the importance of screening in this system. When screening
is accounted through the Thomas-Fermi (TF) theory, the log singularity is 
suppressed and, surprisingly, Lorentz invariance is recovered. 
We show that this result is due to the suppression of the intra-band transitions 
relative to the inter-band transitions. 
In order to test our findings, we study the problem of cyclotron resonance in this system 
and find good quantitative agreement with recent measurements \cite{Henriksen08}. 
\begin{figure}
 \centering
 \includegraphics[width=6cm,angle=-90,bb=91 40 558 719]{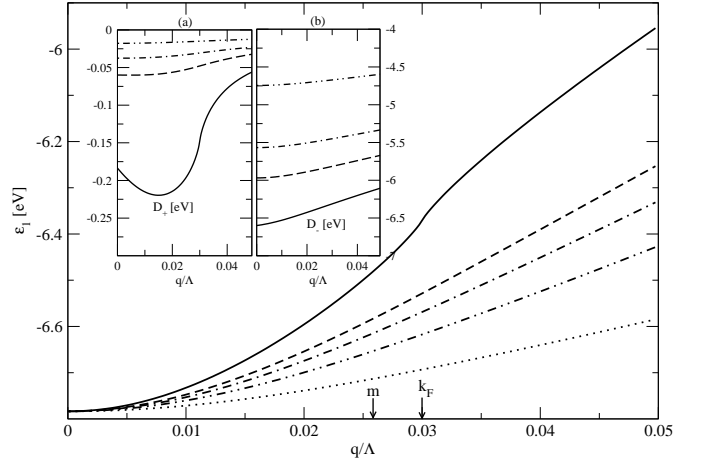}
 \caption{Quasiparticle dispersion, $\e_1$(q) (in eV), as a function of the momentum $q$
 (in units of $\L$). Inset: intra ($D_+$) and inter ($D_-$) band
contributions. Full: $\b=0$ (HF); Dashed: $\b=1$; Dash/dot:$\b=2$; 
Dash/dot-dot: $\b=5$; Dotted line: non-interacting. 
Curves have been shifted for comparison ($k_F/\L=0.03$ and $g=0.5$).}
 \label{fig:UKscr}
\end{figure}
  
We use a tight binding description of BLG in which only the in-plane, $t\approx3$eV, and
the out-of-plane, $t_\perp \approx0.37$eV,  nearest neighbor hopping 
 parameters are considered \cite{note}. In this case we have $\tilde{c} = 3 t a/2$ ($a=1.42 \mbox{\r{A}}$)
 and $m \tilde{c}^2 = t_\perp /2$. From now on we choose units such that
 $\hbar=1=\tilde{c}$. The hyperbolic shape of the
 non-interacting dispersion introduces an intrinsic energy scale in the
 problem, $m$. In the ``non-relativistic'' (NR) 
limit, $k\ll m$, one can replace the four hyperbolic bands by
two parabolic  bands $E_{\pm}(k)=\pm k^2/2m$~\cite{falko} . 
In this approximation, the usual NR dispersion of the 2DEG is
recovered, but with allowed negative energy values. In the ``ultra-relativistic'' (UR)
limit, $k \gg m$, $E(k) \sim k$, one obtains the SLG dispersion. 
The crossover energy scale from non-relativistic to ultra-relativistic
(NR-UR) is given by $m$. We notice that the effective non-relativistic low energy
approximation fails when treating
the interacting problem because the Coulomb energy
associated with electron-electron interactions is of the order of the
inter-band transitions \cite{Silvia08}.

\begin{figure}
 \centering
 \includegraphics[width=6cm,angle=-90,bb=85 49 558 715]{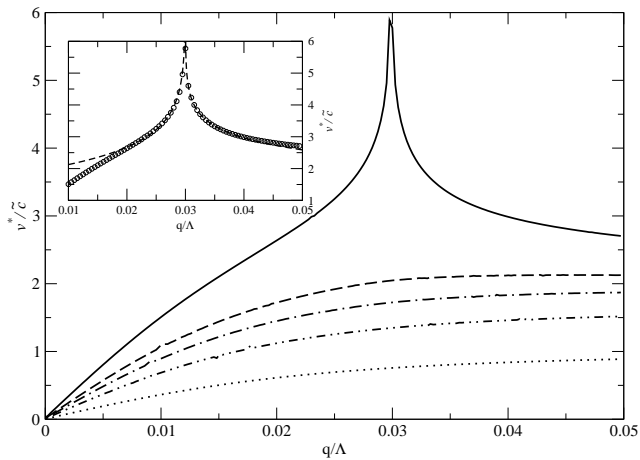}
 \caption{$v^*(q)$ (in units of $\tilde{c}$)
 as a function of $q$ (in units of the cutoff $\L$) for $k_F/\L=0.03$.  
Dotted line: non-interacting. Solid line: HF. Dashed line: TF
with $\b=1$; Dash/dot: $\b=2$; Dash/dot-dot: $\b=5$. Inset: zoom-in near the Fermi vector and 
the logarithmic fit to the divergence (dashed line) for HF. }
 \label{fig:vel}
\end{figure}

The electronic interactions are included by adding to the non-interacting
energy $E_0=1/\pi^2\sum_i\int E_i(p)d\vp$ an exchange term 
which can be written as (energies are given per unit area, and the spin and
valley degeneracy factor of $4$ is accounted):
\begin{eqnarray} \label{Eex1} 
E_{ex} \!\!=\!\!-\!2 \!\!\sum_{i,j\a}\!\! \int_{p,q}
\x_{ij}^\a(\bq,\bp)\x_{ji}^\a(\bp,\bq)n_{i}(q)n_{j}(p)V_\a(\bq-\bp)
\end{eqnarray}
where $\a \!\!=\!\! \pm \!1$ correspond to the symmetric/antisymmetric representations
of the Coulomb interaction:
\begin{eqnarray}
V_{\pm}(\bk)= 2\pi e^2 (1 \pm \exp\{-kd\})/[2\e (k + \beta m)] \,
,
\label{vcou}
\end{eqnarray}
$n_{i}(q)$ 
is the occupation number of band $i$, and $\x_{ij}^\a(\bq,\bp)$ are overlap
matrices which contain information of the change of basis \cite{johan_ferro}. Screening is taken into account through 
the TF approximation by introducing a screening length in
(\ref{vcou}) that is proportional to  the density of states. In (\ref{vcou}) $\b$ is the parameter that 
controls the value of the TF screening length, the HF theory is obtained by taking $\b =0$. Within the Random Phase Approximation (RPA), $\b_{RPA}=4g\l(1+E_F/m\rg)$, being $g=e^2/\hbar \tilde{c}$ the dimensionless coupling constant, and $E_F$ the Fermi energy. For experimentally realized densities ($n_e\approx 10^{11}$ - $10^{13}$ cm$^{-2}$) and $g=0.5$ ($\e=3.9$ for SiO$_2$), $\b \approx$ 1 - 5. 
The energy of a quasiparticle in the $i^{th}$ band is given by 
$\e_i(q)=\d E/\d n_i(\vq)\vert_{n_i=n_i^0}$, where 
$\d n_i(\vq)=n_i(\vq)-n_i^0(\vq)$, being $n_i^0(\vk)$ the occupation number
of the non-interacting system. $E[\d n_i]$ is the total energy $E=E_0+E_{ex}$. 
We can therefore write $\e_i(q)=E_i(q)+\D E_i(q)$ with 
$\D E_i(q)
=-4\int_q \sum_{\a,j}\x_{ij}^\a(\bq,\bp)\x_{ji}^\a(\bp,\bq)n^0_{j}(p)V_\a(\bq-\bp)$
 the correction to the non-interacting band $E_i(q)$.

We consider the case of electron doping such as that the chemical potential
does not reach the uppermost band, which is usually the 
experimentally realized situation \cite{note2}. 
Therefore, our results are valid for Fermi energies up to 
$\sqrt{2}t_\perp$, which corresponds to densities smaller than 
$n_e\approx 10^{13}$ cm$^{-2}$. We look then at the 
correction to the first band, which we write as $\D E_1(q)=D_+(q,k_F)+D_-(q,\L)$ to distinguish intra-band ($D_+$) from inter-band ($D_-$) contributions. The expressions for $D_\pm$ can be easily derived from $\D E_1(q)$. $k_F$ is the Fermi wave vector and $\L$ a cut-off of the order of the inverse lattice spacing ($\L\approx 1
\mbox{\AA}^{-1}\approx 7$eV). 

\begin{figure}
 \centering
 \includegraphics[width=6cm,angle=-90,bb=91 46 566 754]{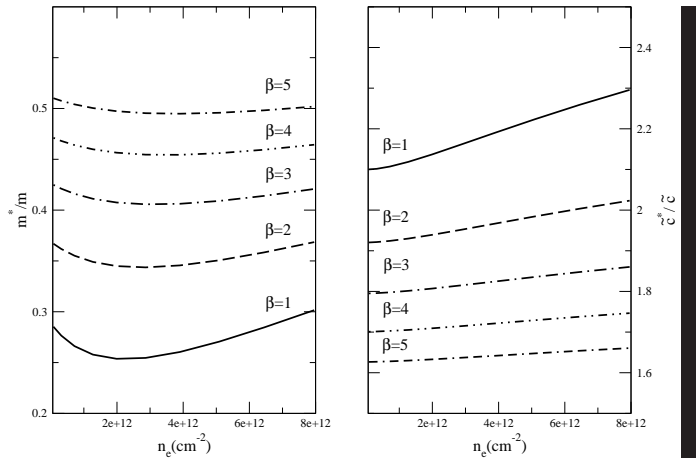}
 \caption{Left: $m^*$ (in units of the bare mass $m$)
as a function of electron density $n_e$ (in units of electrons per cm$^{-2}$).
Right: $\tilde{c}^*$ (in units of the bare velocity $\tilde{c}$) as a function of $n_e$, 
for different $\b$ ($g=0.5$).}
 \label{fig:MandV}
\end{figure}

Fig. \ref{fig:UKscr} shows the quasiparticle band within the HF theory
(solid line) for a typical value of the
Fermi vector.
Fig. \ref{fig:UKscr} (a) depicts the correction due to the intra-band
transitions, $D_+$. Its behavior, as expected, is qualitatively very similar 
to that of a 2DEG \cite{vignale}. In particular, the inflection point seen at 
$q \approx k_F$ is due to the special role of $k_F$ which separates a domain 
with an avoidable singularity $q\leq k_F$ from a singularity free domain for
$q > k_F$. While $D_+$ diminishes with $k_F$, the correction due to $D_-$ is independent of it. The latter is shown in Fig. \ref{fig:UKscr} (b). As it can be seen from the figure, for typical electronic densities,
the correction due to inter-band interactions is roughly two orders of 
magnitude bigger than that of the intra-band. Notice, from Fig.~\ref{fig:UKscr}, 
that the quasiparticle dispersion however inherits the inflection point 
from $D_+$ at $q=k_F$.
The renormalized band velocity is given by $v^*(q)=\vert\partial\e/\partial
q\vert$, which is plotted in Fig. \ref{fig:vel}. Due to the sharp inflection point in $\D E_1$
at $k_F$, the effective quasiparticle Fermi velocity, 
$v^*(k_F)$ presents an unphysical logarithmic divergence: $v^*(k \sim k_F) \sim -4g/\pi\log(\vert k- k_F\vert/\L)$, as it occurs for the 2DEG. For small momentum nevertheless, $q/k_F\ll 1$, the renormalized dispersion can be shown to be 
parabolic: $\e_1(q)\approx q^2/(2\tilde{m})$, with $\tilde{m}^{-1}=m^{-1}+ g (5/k_F-1)/2$.

\begin{figure}
 \centering
 \includegraphics[width=6cm,angle=-90,bb=87 39 566 754]{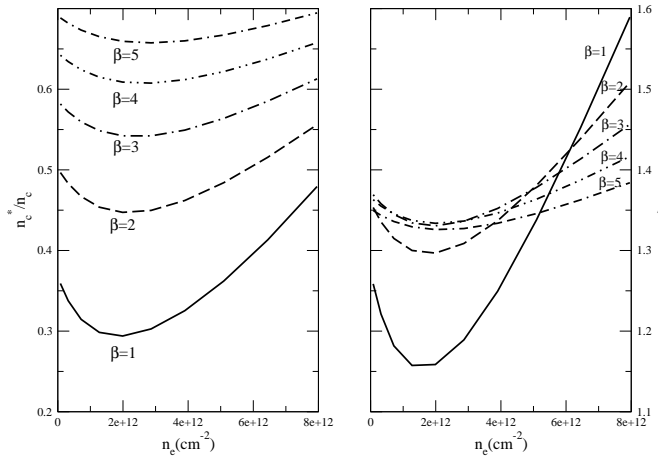}
 \caption{Left: NR-UR crossover density $n_c^*$ (in units of the bare crossover
 density $n_c$) as a function of electronic density $n_e$ (in units of
 electrons per cm$^{-2}$). Right: $E_c^*$ (in units of the non-interacting
 crossover energy $E_c$) as a function of $n_e$, for different $\b$ ($g=0.5$).}
 \label{fig:BandE}
\end{figure}

As mentioned earlier, the divergence of the Fermi velocity is an
unpleasant feature of the HF approximation which indicates the necessity of 
introducing screening in the problem.
The renormalized band $\e_1(q)$ is shown in Fig. \ref{fig:UKscr} for
different values of $\b$. We see that the introduction of
screening eliminates the inflection point at $k_F$. This can be seen clearly
in Fig. \ref{fig:vel}, where it is shown that the 
divergence in the quasiparticle velocity disappears for finite $\b$. 
The most striking feature of our calculations is that the quasiparticle 
dispersion can be fitted by a non-interacting like dispersion:
\begin{eqnarray}
\e_1(k)=\e_0+\sqrt{(v^*_F k)^2+(m^*\tilde{c}^{*2})^2} \, ,
\label{quasi}
\end{eqnarray} 
where $\e_0$ is a constant, and $m^*$ and $\tilde{c}^*$ are the quasiparticle mass
and renormalized ``light'' velocity, respectively.
We find that this result is valid to high accuracy for a large region of
energy and momenta due to the fact that
the inter-band transitions largely dominate over the intra-band ones \cite{note3}.

The results for $m^*/m$ and $\tilde{c}^*/\tilde{c}$ are shown in Fig. \ref{fig:MandV} as
a function of electronic 
density, $n_e=k_F^2/\pi$, for different values of the screening strength $\b$. 
Note that, for fixed $\b$, $\tilde{c}^*/\tilde{c}$ increases monotonically with density, whereas $m^*/m$ 
has a minimum at a finite $n_e$. While $m^*/m$ is renormalized 
to smaller values, $\tilde{c}^*/\tilde{c}$ is renormalized to larger values. 
This has interesting consequences for the NR-UR crossover mentioned earlier. 
The crossover energy for the non-interacting problem is given by $E_c=m
\tilde{c}^2$. Analogously, for the interacting result we 
can define the crossover energy as $E_c^*=m^* (\tilde{c}^*)^2$. 
This quantity is plotted in Fig. \ref{fig:BandE} (b), which shows that 
$E_c^*> E_c$ for all the values of the parameters (this is also true if we 
vary the coupling constant $0.1\leq g\leq 2$). 
However, the relevant parameter to compare with experiment is the crossover 
electronic density, $n_c=q_c^2/\pi=(m\tilde{c})^2/\pi$, that is, the density at
which the NR-UR crossover takes place. Fig. \ref{fig:BandE} (a) shows the 
renormalized value of this quantity, $n_c^*=(m^*\tilde{c}^*)^2/\pi$, in units of the
 non-interacting value $n_c$. Indeed, it is seen that $n_c^*<n_c$ always, 
even though the renormalized quasiparticles' energy is higher.

Let us now consider the problem in the presence of a transverse magnetic
field $B$. For the non-interacting problem, the Landau levels
are given by  (restoring units) \cite{Guinea06}:
\begin{eqnarray} \label{ELL}
\frac{E^{\pm}_n}{\omega_c} \!\!=\!\! \pm  
\left\{\! n\!+\! \frac{1}{2} \!+\!  2 r^2
\!\!-\!\! \frac{1}{2} \! \left[1 \!+\! 16 r^4 \!\!+\! 16 r^2 
\left(\!n\!+\! \frac{1}{2}\!\right)\right]^{1/2}\right\}^{1/2}
\end{eqnarray}
where $n$ is a positive integer, $\omega_c =\tilde{c}\sqrt{2eB/c}$ is the
cyclotron frequency, and $r=m v^2_F/\omega_c$. 
One can clearly see that this problem has the NR-UR crossover as a function of $B$
discussed earlier. At low fields, $r \gg 1$, 
we find $E^{\pm}_n \approx \pm  [\omega_c^2/(2 m \tilde{c}^2)] \, \sqrt{n (n+1)}$, 
and the Landau level energy is proportional to $B$ as in the NR
problem \cite{falko}; 
at high fields, $r \ll 1$, one finds $E^{\pm}_n \approx \pm \omega_c \, \sqrt{n}$ and, as in the UR case, we find the Landau level
energy proportional to $\sqrt{B}$.  

\begin{figure}
 \centering
 \includegraphics[width=6cm,angle=-90,bb=86 36 558 709]{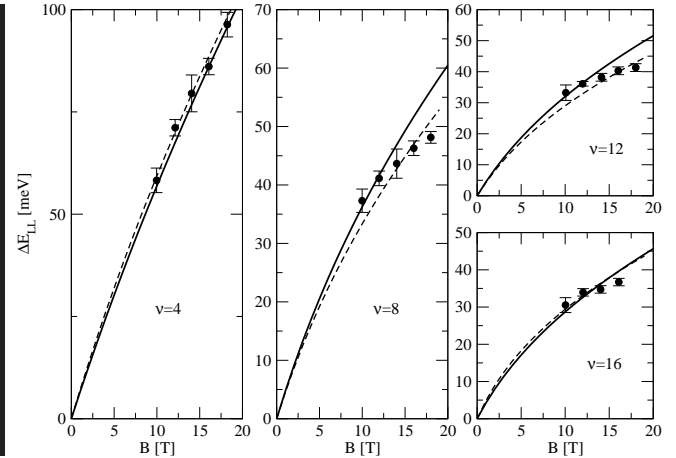}
 \caption{Landau level transition energies for different filling factors
 $\nu=4$ ($n$: $0\rightarrow 1$), $\nu=8$ ($n$: $1\rightarrow 2$), $\nu=12$ 
($n$: $2\rightarrow 3$) and $\nu=12$ ($n$: $2\rightarrow 3$) as a function of
$B$ (in Tesla). The experimental data (dots) was taken from
Ref. \cite{Henriksen08}. The solid line is the theoretical value for
$g=0.5$, and $\b=4$ for bare values $\tilde{c} \approx 0.76 \times 10^6$ m/s and
$t_{\perp} \approx 0.33$ eV. The dashed line shows the fits obtained in 
Ref. \cite{Henriksen08} for $t_\perp=0.35$ eV.} 
 \label{fig:ELLfit}
\end{figure}

Just as in the case of a Fermi liquid, here the quasiparticles 
carry electric charge $e$ and  
couple to a magnetic field via minimal coupling. Note that here, however, the cyclotron mass is not protected by Kohn's theorem since the dispersion is not parabolic \cite{Kohn61,note4}. Hence, the Landau level spectrum is the same as the non-interacting problem, eq. (\ref{ELL}),
with the bare parameters, $m$ and $\tilde{c}$, replaced by renormalized ones, $m^*$ and $\tilde{c}^*$, respectively. In Fig. \ref{fig:ELLfit} we show
the data from cyclotron resonance experiments \cite{Henriksen08} for
inter-Landau level transitions for different filling factors $\nu =
n_e/n_{\Phi}$ ($n_{\Phi}$ is the density of flux quanta through the system) 
together with our results for $g=0.5$,
$\beta=4$, $\tilde{c} = 0.76 \times 10^6$ m/s ($\tilde{c}^*=1.2 - 1.3 \times 10^6$ m/s) and
$t_{\perp} \approx 0.33$ eV ($t_{\perp}^*=0.44 - 0.45 $ eV). These values were obtained by fitting our renormalized theory to the experimental data, taking $\beta$, $g$, and the bare $m$ and  $\tilde{c}$ as free parameters, and are in agreement with recent infrared spectroscopy data \cite{ZQLi08}. The variation of $\tilde{c}^*$ and $m^*$ with electronic density was taken into account, however we took the screening strength $\beta$ as fixed, since our treatment is not self consistent. Nevertheless, as a double check, the value obtained for $g$ is the expected for BLG on SiO$_2$ and $\beta$ falls within the expected range for such densities.
One can see that our results are in fair quantitative agreement with the experimental data, giving support to
the idea that this system can be described by a Dirac liquid of quasiparticles with a dispersion given by (\ref{quasi}). There is a small electron-hole asymmetry due to inter-band interactions (and therefore independent of density), which results in smaller values of $t_{\perp}^*$ ($\sim10\%$) and $\tilde{c}^*$ ($\sim5\%$) for hole doping. This difference however is not enough to explain the asymmetry observed in \cite{Henriksen08}. As it was mentioned above, the screening strength $\b$ was fitted to a constant value for all four plots in Fig. \ref{fig:ELLfit}, and it should be taken as the best average $\b$ that gives a reasonable good fit for all the data range. For a better agreement with the data it would be probably necessary to include selfconsistently the dependence of $\b$ with the electronic density but this goes beyond the scope of the present paper.

We have studied the effect of electron-electron interactions on the
electronic properties of a graphene bilayer within
the Hatree-Fock-Thomas-Fermi theory by taking into account the full four-bands model of BLG. 
We have shown that the quasiparticles can be described by a non-interacting, Lorentz like dispersion with renormalized parameters which depend on the electronic density. The fact that the Lorentz invariance of the dispersion is recovered for a large range of energies is an unexpected result since there is no evident symmetry behind it. It is important to note that this result is due to the dominance of inter-band transitions, which are missing in the usual Fermi liquid picture. Since this contribution is independent of electronic density, the accuracy of the effective description increases with decreasing density. Further investigation is needed to determine if corrections beyond HF lead to deviations from the Lorentz dispersion. Furthermore, we have tested our
calculations by comparing our results with recent cyclotron resonance
experiments and found quantitative agreement between theory
and experiment. Our results are also in agreement with recent {\it ab-initio} calculations \cite{Gruneis08}.

The authors wish to thank A. K. Geim, Z. Q. Li, A.H. MacDonald and Johan Nilsson for their valuable comments on the manuscript.



\begin{thebibliography}{99}
\bibitem{note0} 
Note that the mass renormalization in both the non-relativistic and relativistic Fermi liquid is directly associated to the Galilean or Lorentz invariance of the system, see Refs. \cite{baym,pethick}.
\bibitem{note} Other hopping parameters are smaller in magnitude \cite{gic}
and usually suppressed by disorder \cite{johan}.
\bibitem{note2} The case of hole doping is analogous, except for a small asymmetric term \cite{Silvia08} which we discuss later.
\bibitem{note3} Note that the quasiparticle Fermi velocity, $v^*(k_F)$ is not inversely proportional to the effective mass (as in the usual Fermi liquid case), but to the quasiparticle energy.
\bibitem{note4}
Kohn's theorem might be expected to be asymptotically valid in the NR limit. However, given the renormalization of the UR-NR crossover to lower electronic densities, this would be the case only at very low densities. In the dilute limit inhomogeneities \cite{yacoby}, Wigner crystallization \cite{Silvia08,dahal07} and ultimately the trigonal warping term \cite{falko} would also be relevant.
\end{thebibliography}
\end{document}